# Decoration of growth sector boundaries with single nitrogen vacancy centres in as-grown single crystal HPHT synthetic diamond

P. L. Diggle[1,2], U. F. S. D'Haenens-Johansson[3], B. L. Green[1],
C. M. Welbourn[1], Thu Nhi Tran Thi[4], A. Katrusha[5], W. Wang[3], and M. E. Newton[1,2,*]

[1] *Department of Physics, University of Warwick, Coventry, CV4 7AL, UK.*
[2] *EPSRC Centre for Doctoral Training in Diamond Science and Technology, UK.*
[3] *Gemological Institute of America, New York City, NY, USA.*
[4] *European Synchrotron Radiation Facility (ESRF), Grenoble, France*
[5] *New Diamond Technology Ltd, St. Petersburg, Russia*

**Abstract**

Large (> 100 mm$^3$), relatively pure (type II) and low birefringence single crystal diamond can be produced by high pressure high temperature (HPHT) synthesis. In this study we examine a HPHT sample of good crystalline perfection, containing less than 1 ppb (part per billion carbon atoms) of boron impurity atoms in the {001} growth sector and only tens of ppb of nitrogen impurity atoms. It is shown that the boundaries between {111} and {113} growth sectors are decorated by negatively charged nitrogen vacancy centres (NV$^-$): no decoration is observed at any other type of growth sector interface. This decoration can be used to calculated the relative {111} and {113} growth rates. The bulk (001) sector contains concentrations of luminescent point defects (excited with 488 and 532 nm wavelengths) below $10^{11}$ cm$^{-3}$ ($10^{-3}$ ppb). We observe the negatively charged silicon-vacancy (SiV$^-$) defect in the bulk {111} sectors along with a zero phonon line emission associated with a nickel defect at 884 nm (1.40 eV). No preferential orientation is seen for either NV$^-$ or SiV$^-$ defects, but the nickel related defect is oriented with its trigonal axis along the $\langle 111 \rangle$ sector growth direction. Since the NV$^-$ defect is expected to readily re-orientate at HPHT diamond growth temperatures, no preferential orientation is expected for this defect but the lack of preferential orientation of SiV$^-$ in {111} sectors is not explained.

---

[*] Corresponding Author. University of Warwick, CV4 7AL, UK. Email address; m.e.newton@warwick.ac.uk

# 1   Introduction

Two methods are routinely employed for the synthesis of single crystal diamond: high pressure, high temperature (HPHT) synthesis or chemical vapour deposition (CVD). Single crystal HPHT diamonds are grown on a small seed crystal using the temperature gradient method in the diamond thermodynamic stability region. Typical solvents are alloys of iron, nickel, and cobalt. Nitrogen has a high solubility in HPHT grown diamond, and can act to change the morphology and/or growth rate of the diamond crystal.[1,2] Nitrogen incorporation can be reduced by the addition of a nitrogen "getter" to the metal solvent, typically titanium, aluminium, or zirconium which acts to form a stable insoluble complex with nitrogen.[3] In low nitrogen HPHT diamond, boron is often the dominant impurity due to it being present in the carbon source, the solvent/catalyst[4] or even the HPHT capsule materials. It is well understood that the morphology of the HPHT diamond is largely governed by the relative development of {001} and {111} sectors as controlled by the temperature and pressure[5] but also the presence of impurities. The {011}, {113} and {115} sectors are often present when the crystal is grown with nitrogen getters to produce type IIa (nitrogen impurities typically less than ~1 ppm) diamond,[3] and are also seen with the addition of boron to the source material for the production of type IIb (neutral boron impurities detectable by FTIR absorption) diamond.[6] The uptake of impurities is growth-sector dependent, as summarised by the modified Kanda diagram.[7]

In CVD synthesis, the thermal dissociation of a carbon-containing gas and hydrogen, followed by carbon deposition, yields epitaxial growth of a diamond crystal at below atmospheric pressure.[8] For CVD growth, the chemical purity of the synthesis environment can be controlled with the precision of a semiconductor growth process and hence this approach has undergone a vast research effort in recent years. CVD grown diamond now facilitates a whole host of applications ranging from optical and radiation windows[9–11] to quantum technologies[12,13]. However, keeping the impurity concentrations (such as nitrogen and boron) under 1 part per billion (ppb, $10^{14}$ cm$^{-3}$) whilst simultaneously reducing the concentration of dislocations through the material is a challenge.[14] Strain arising from dislocations, and induced birefringence, can limit the material performance in optical, phononic, and quantum technology applications.

CVD synthesis is further limited in its ability to produce single crystal diamond of large area, with high purity, and low strain. For homoepitaxial growth, the size of the diamond substrate is the limiting factor: tiling (forming a mosaic) of substrates has been demonstrated but generates a high concentration of dislocations at the substrate interfaces.[15] Heteroepitaxial growth is showing significant advances toward the production of large single crystal samples, but is still plagued with high (though relatively uniform) dislocation densities.[16]

Given the significant control of point defect incorporation through CVD growth, the recent advancements in HPHT diamond synthesis have been somewhat overshadowed. The control of the HPHT diamond growth chemistry has improved such that large diameter (> 10 mm) high purity diamond can be grown.[4,17] The ability to grow large HPHT material with very low dislocation densities is potentially enabling for several technological areas.[4,17–20] However, in contrast with CVD growth where the challenge for high-purity high-crystalline quality material lies primarily with the reduction of the density of dislocations, for HPHT synthesis it is still a challenge to produce diamond with concentrations of nitrogen and boron impurities of less than 50 ppb.

The drive toward better understanding of diamond synthesis and its subsequent improvement is partially underpinned by the potential of diamond-based applications utilising point defects as quantum systems in which spin states can be optically initialised, manipulated, and read out at room temperature.[21] One such point defect in diamond is the heavily-studied negatively charged nitrogen vacancy centre ($NV^-$: Substitutional nitrogen impurity adjacent to a lattice vacancy with a trapped electron) which can be created during-[22,23] or post-growth.[24–26] The $NV^-$ centre has helped establish diamond as a viable host material for quantum technologies[21,27] including highly sensitive detection of magnetic fields, electric fields, temperature, pressure, as well as communication and computation.[28–33]

In addition to potential technological exploitation, the presence and properties of grown-in point defects can be instrumental in understanding the growth and impurity incorporation mechanisms during synthesis. For example preferential orientation of the defect symmetry axis with respect to the growth direction of $NV^-$ centres was first shown in CVD grown diamond on a {110} substrate,[34] whilst optically detected magnetic resonance (ODMR) measurements further demonstrated this for CVD diamond grown on {111} or {113} substrates,[22,23] indicating that the defect incorporates as a unit (nitrogen first with a vacancy in the layer above) rather than being produced by the capture of a mobile vacancy at a substitutional nitrogen impurity.

In this paper we use a variety of well-established techniques, including diffraction limited scanning confocal photoluminescence (PL) microscopy, to investigate the incorporation of point and extended defects in a HPHT diamond sample.

## 2    Experimental Details

A $4.0 \times 4.0 \times 0.5$ mm plate was cut from a HPHT diamond sample produced by New Diamond Technology Ltd (NDT) with the large face being within $1^o$ of the (001) plane as confirmed by Laue X-ray diffraction. This HPHT diamond was synthesised in a Co-Fe-C system, using a proprietary nitrogen getter. The sample had not been treated in any way (i.e. annealing, irradiation etc.) post growth. Electron paramagnetic resonance (EPR) measurements, a bulk technique averaging over the whole sample, was used to measure the mean neutral substitutional nitrogen concentration.

For low temperature cathodoluminescence (CL) imaging and spectroscopy, measurements were carried out using a Scanning Electron Microscope (Zeiss EVO 50) equipped with a bespoke low magnification CL imaging system (PMT, Hamamatsu R374) and a Gatan MonoCL3 system for UV and visible spectroscopy measurements (PMT, Hamamatsu R943-01, CCD, Princeton Instruments Pixis 100). The sample was sputtered with a thin layer of gold to avoid any charging effects, and mounted onto a low temperature stage (Gatan C1002) and held at liquid nitrogen temperatures (77 K).

Confocal PL imaging was conducted with a home-built room temperature setup using 488 and 532 nm excitation equipped with an oil objective (Zeiss Plan Apo 100x, NA = 1.4). The sample was mounted onto a co-planar waveguide housed on top of a 3-axis piezoelectric stage. Fluorescence was detected on two single photon counters (Excelitas SPCM-NIR), and filtered by an appropriate Semrock transmission filter depending upon the point defect to be imaged. A 50:50 beamsplitter partitioned the fluorescence to both detectors, enabling second order photon autocorrelation measurements to be conducted. PL spectra were taken by a fibre coupled spectrometer (ANDOR SR-303i-B). For ODMR measurements, a 20 μm copper wire was placed over the top of the sample for microwave delivery and an external magnetic field applied to the sample via a permanent magnet mounted on a manual 5-axis stage.

## 3 Results

### 3.1 Material Quality

The material quality is assessed by means of white light cross polarisation imaging and white beam X-ray topography (XRT) which show strain and extended defects respectively. We note that this sample contains a small metallic inclusion and a small fracture in the crystal, as shown in Figure 1. Furthermore, the material shows low internal strain where the average dislocation density is below $10^3$ cm$^{-2}$, and much lower in (001) growth sectors. XRT reveals stacking faults in the {111} growth sectors.

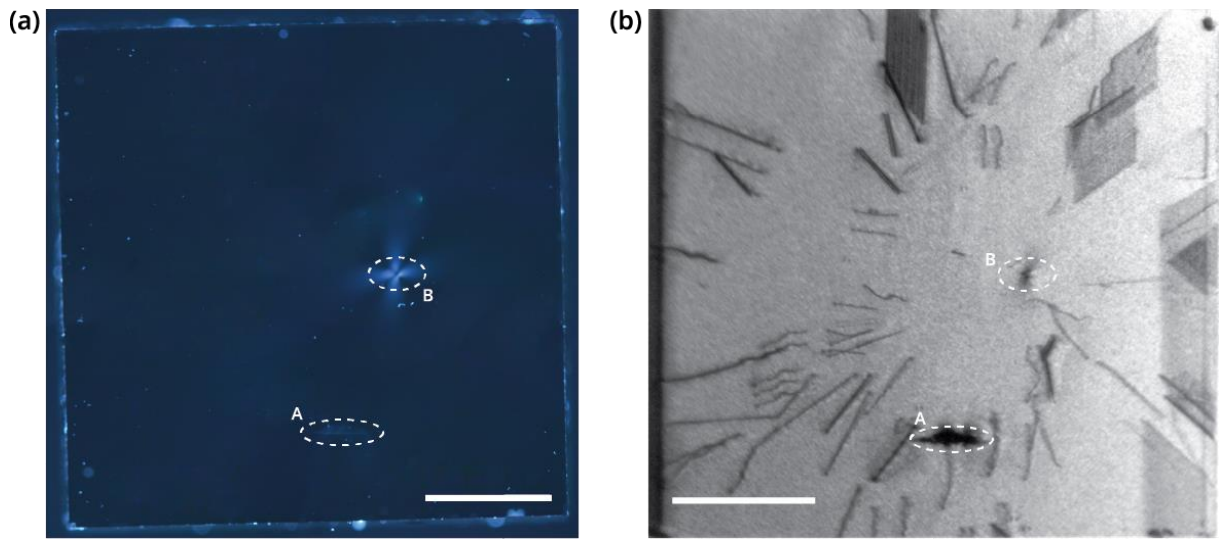

*Figure 1 (a) Cross polarisation image of the NDT HPHT sample (polars vertical and horizontal) showing a small fracture in the crystal (A) and an inclusion (B). (b) White beam X-ray topograph (400 reflection) with the fracture and inclusion highlighted. (Scale bar: 1 mm).*

### 3.2 Impurity and point defect incorporation

EPR measurements indicated that the average concentration of substitutional nitrogen [$N_s^0$] in the sample studied is 6.5 ± 1 ppb. This indicates that in certain growth sectors, at least, the concentration of substitutional nitrogen is higher than the concentration of substitutional boron. The positively-charged substitutional nitrogen concentration ($N_s^+$), as produced by the presence of both the substitutional boron ($B_s$) acceptor and nitrogen, where $N_s^0 + B_s^0 \rightleftharpoons N_s^+ + B_s^-$, was below infrared detection limits.

Visible CL imaging of the sample, Figure 2, shows cuboctahedral growth, with many minor sectors. In Figure 2 a model (001) slice taken from a cuboctahedral crystal displaying {001},

{011}, {115}, {113} and {111} sectors is compared to the CL image of the sample. The boron bound exciton (BE) signal, relative to the free exciton (FE), observed in low temperature (77 K) CL spectroscopy is used to determine the concentration of substitutional boron impurities.[35,36] The luminescence intensity ratio between the BE and FE is considered proportional to the total shallow acceptor (or donor) concentration, independent of the compensation state of the semiconductor where all shallow impurities are efficiently neutralized under electron injection.[37] In other work a comparison between CL and secondary ion mass spectroscopy (SIMS) indicates that in the as grown samples all incorporated boron is in the form of substitutional boron.[36] In the system used here, we estimate a detection limit < 1 ppb for the substitutional boron acceptor. The boron BE is detected in all growth sectors, except in the {001} sector. The average substitutional boron concentration in each growth sector is given in Table 1.

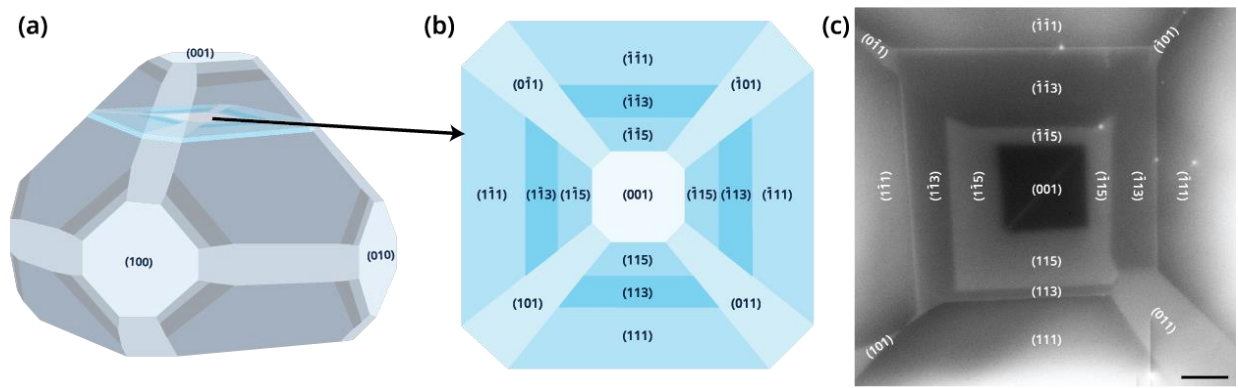

*Figure 2: (a) Representation of an idealised cuboctahedral diamond crystal. (b) A (001) plate taken from the position shown in (a), with the major and minor growth sectors labelled. (c) Low temperature (77 K) cathodoluminescence image (visible light emitted) of the NDT HPHT sample studied. (Scale bar: 200 μm).*

*Table 1: Average substitutional boron acceptor concentration (ppb) in the growth sectors as determined by low temperature (77 K) CL spectroscopy.*

|  | Growth sector | | | | |
| --- | --- | --- | --- | --- | --- |
|  | (001) | {011} | {115} | {113} | {111} |
| **[B] (ppb)** | < 1 | 19 ± 3 | 2 ± 1 | 9 ± 2 | 84 ± 10 |

The survey of the (001) sector does not show any room temperature PL emission associated with defects that are excited by 488 or 532 nm wavelengths. Since the detection limit of the scanning confocal microscope is on the order of $10^{11}$ cm$^{-3}$ ($10^{-3}$ ppb) for common luminescent point defects in diamond this implies that virtually no (or very few) luminescent point defects are within this sector. By translating the sample and moving from {001} into {115} and then into {113}, no point defect emission was observed. However, when the sample was translated towards the {111} growth sectors, a locally-high concentration of NV$^-$ centres was observed at the {111} to {113} growth sector boundary.

Figure 3 highlights one corner of the sample, comparing the same region of the sample with CL and confocal PL imaging. Here it was possible to observe several growth sector interfaces within a 300 μm x 300 μm area, and we find that only the {111} to {113} interfaces exhibit fluorescence associated with NV$^-$. The variation in decorations at these interfaces are correlated with the CL image such that a wider region of contrast between certain {111} and {113} sectors results in the appearance of "tram lines" when mapping the NV$^-$ PL (Figure 3, location B & C).

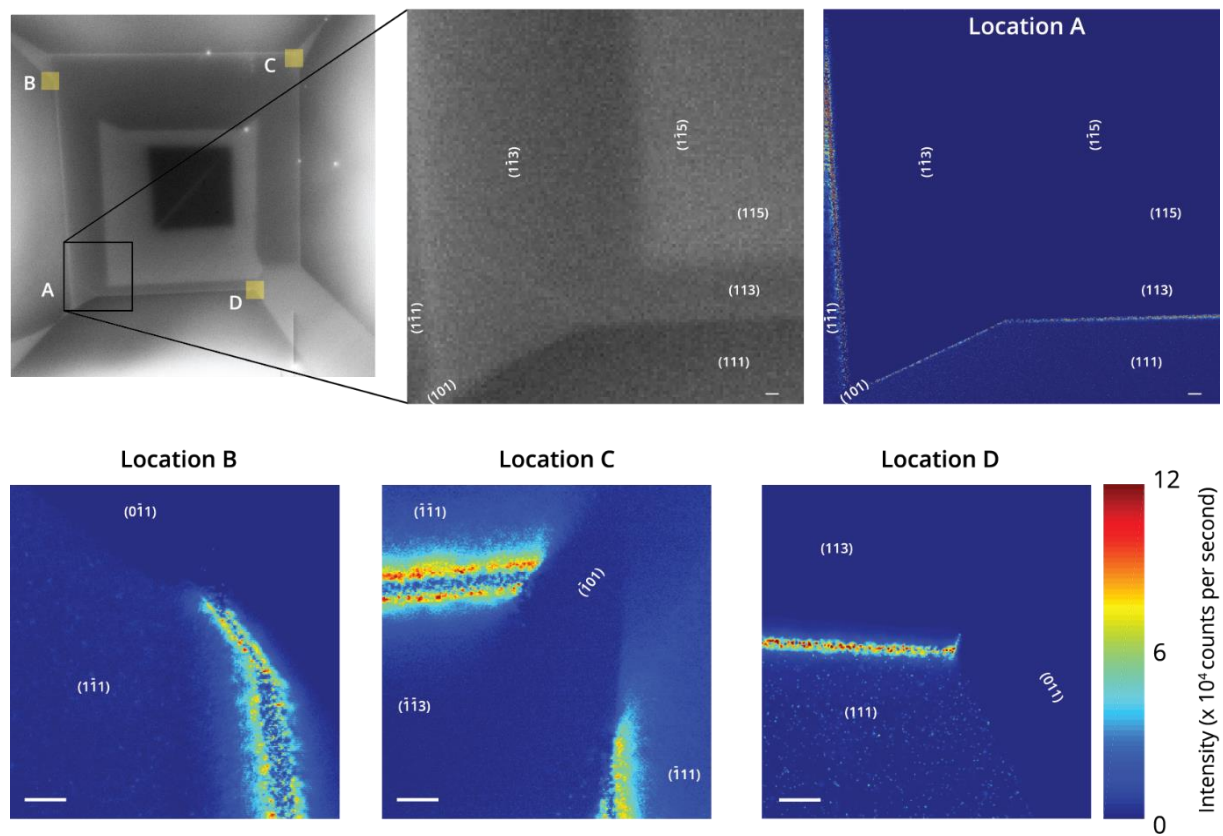

*Figure 3: Top; 300 μm x 300 μm area zoomed in from the CL image with the 300 μm x 300 μm confocal image on the right of the same area. Decoration is only seen at the {111}/{113} interfaces with fluorescence only from NV$^-$ at the sector boundaries. Bottom; NV$^-$ incorporation at the other intersections involving {111}, {113} and {011} for locations as indicated on the top CL image. (Scale bar; 10 μm)*

To investigate the evolution of these decorated interfaces, "XZ maps" were measured at the two locations over 15 μm of sample depth. As shown in Figure 4 we see that the NV⁻ decoration was restricted to narrow regions that follow the sector boundaries and without incorporation of NV⁻ in to the bulk growth sectors.

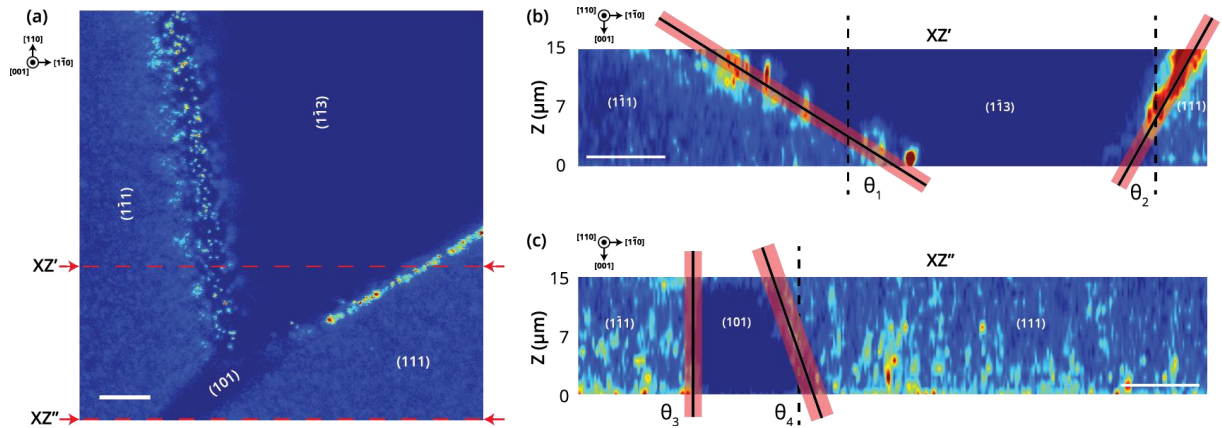

*Figure 4:(a) 80 μm x 80 μm confocal image at a depth of 7 μm from the surface of the sample at location A [Figure 3]. Two locations are marked for the XZ slices. (b) and (c) the resultant planar slices for XZ' and XZ'' respectively. The angles labelled are between the vertical ([001] direction) and the line shown: $\theta_1 = 55°  ± 5°$, $\theta_2 = 30° ± 5°$, $\theta_3 = 0° ± 5°$ and $\theta_4 = 19° ± 10°$. (Scale bar; 10 μm).*

Spectra collected around these interfaces identify three different point defects: NV⁻; SiV⁻ (PL emission at 737 nm); and PL emission centred on 884 nm which is attributed to a nickel-related defect[38,39] labelled previously as the 1.40 eV defect[40] and NIRIM-2[41] and is known to be a paramagnetic (S=1/2) defect of trigonal ($C_{3v}$) symmetry.[38,42] Whilst there are many nickel related defects identified in diamond, a survey of the literature on the 1.40 eV defect highlights that the atomic structure has not yet been definitively proven and so we will continue to refer to it as the 1.40 eV defect from this point forward.

The spatial distribution of each defect close to the {113} and {111} interface can be identified by luminescence mapping using optical filters matched to the emission range of each defect (Figure 5). These maps reveal that the incorporation of defects is inhomogeneous, such that the boundary between {111} and {113} growth sectors is decorated only with NV⁻ centres, whereas only the bulk {111} sectors contain both SiV⁻ and 1.40 eV defects.[40,43] It should be noted that the filters are not 100% selective for each defect, for example the SiV⁻ filter will also pass approximately 10% of the emission from the vibronic side band of NV⁻ defects.

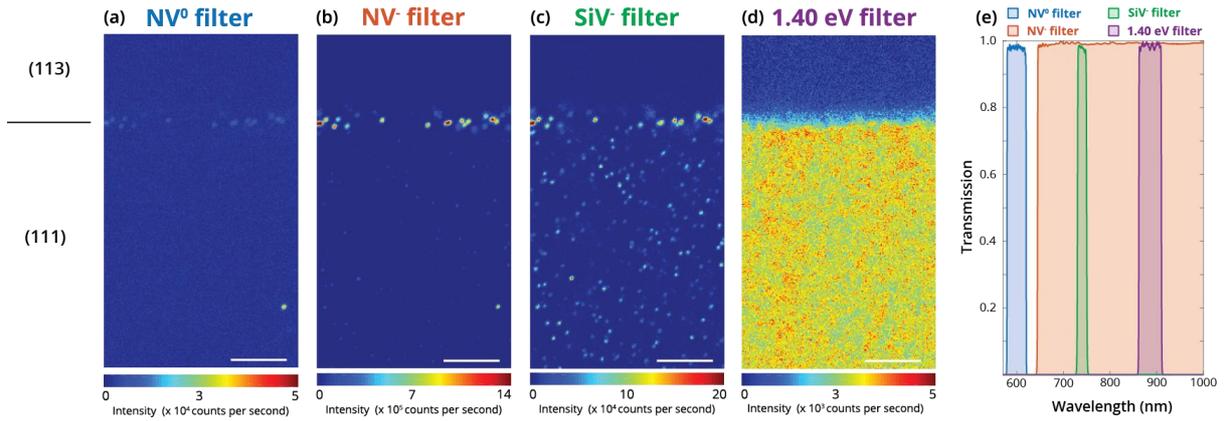

*Figure 5: Employing optical filters matched to the emission range of each defect, the spatial distribution of each defect at the {113} to {111} sector interfaced can be mapped. (a) $NV^0$, (b) $NV^-$ (c) $SiV^-$ (with ~10% $NV^-$) and (d) 1.40 eV defect. The transmission profile of each filter is given in (e). For the 1.40 eV defect shown in (d), the 532 nm excitation power was increased to 100 mW c.f. 5 mW for the other defect distribution maps in this figure. (Scale bar; 5 μm).*

### 3.3 Orientation of incorporated defects

The number of defects within a diffraction-limited optical volume can be quantified by performing second order photon autocorrelation measurements ($g^2$),[24] where the resolution of single point defects in diamond implies a bulk defect concentration on the order of $1.76 \times 10^{11}$ cm$^{-3}$ ($10^{-3}$ ppb) or less. The $g^2$ measurements at several interfaces found that where the $NV^-$ decoration is wider than ~3 μm, the $NV^-$ centres are incorporated primarily as single centres per diffraction-limited volume; otherwise we observe ensembles of $NV^-$ defects where the local concentration of $NV^-$ centres is higher.

For a {111} to {113} interface with an $NV^-$ decoration width of ~5 μm, 78% of the NVs present are single centres (Figure 6). It is possible to identify the orientation of a given $NV^-$ centre by applying an arbitrarily-oriented magnetic field and measuring the optically detected magnetic resonance (ODMR) spectrum: this will identify if there is any preferential (e.g. not random with respect to the possible symmetry related possible orientations) defect alignment.[22,23] Measurements on 100 $NV^-$ centres revealed all four possible orientations occurred with equal probability (0.25) indicating no preferential orientation/alignment.

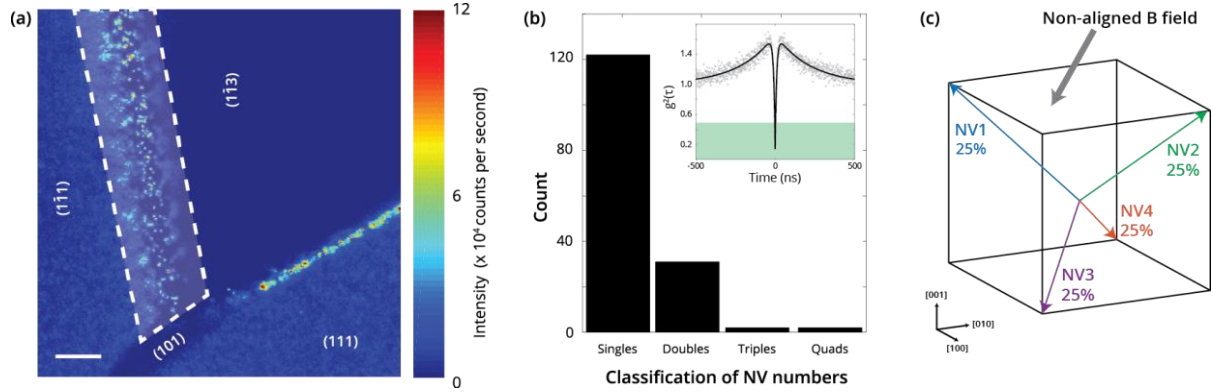

*Figure 6: (a) Optically detected magnetic resonance (ODMR) study at location A [Figure 3] (Scale bar; 10 μm). (b) classification histogram of NV⁻ centres based on their photon autocorrelation minima with a representative measurement for a single centre (inset). (c) The four possible trigonal orientations along ⟨111⟩ for the NV⁻ centre with the measurement occurrence from ODMR in this study.*

Room temperature ODMR on SiV⁻ and the 1.40 eV defect have not been previously reported, therefore we investigated possible preferential alignment of both defects by a rotation of the linear excitation polarisation. The 737 nm ZPL of SiV⁻ is an E ↔ E optical transition[44] (ignoring the small spin-orbit splitting) and the 1.40 eV defect 884 nm ZPL is an A ↔ E optical transition.[42] Both defects are trigonal symmetry defects, and hence the rotation of the linear excitation polarisation will produce a fluorescence intensity modulation which is well understood.[44–48]

Optical polarisation measurements were conducted by rotating the electric field vector of the excitation laser from $E||[1\bar{1}0]$ (horizontal polarisation) to $E||[110]$ (vertical polarisation). SiV⁻ showed equal contrast in two {111} sectors indicating a non-preferential alignment with respect to the growth direction (Figure 7). However, the 1.40 eV defect responds differently such that one sector favoured maximal emission by vertical polarisation and the other sector by horizontal polarisation. We note that the concentration of the 1.40 eV defects in the two {111} sectors is different, as judged by the maximum emission.

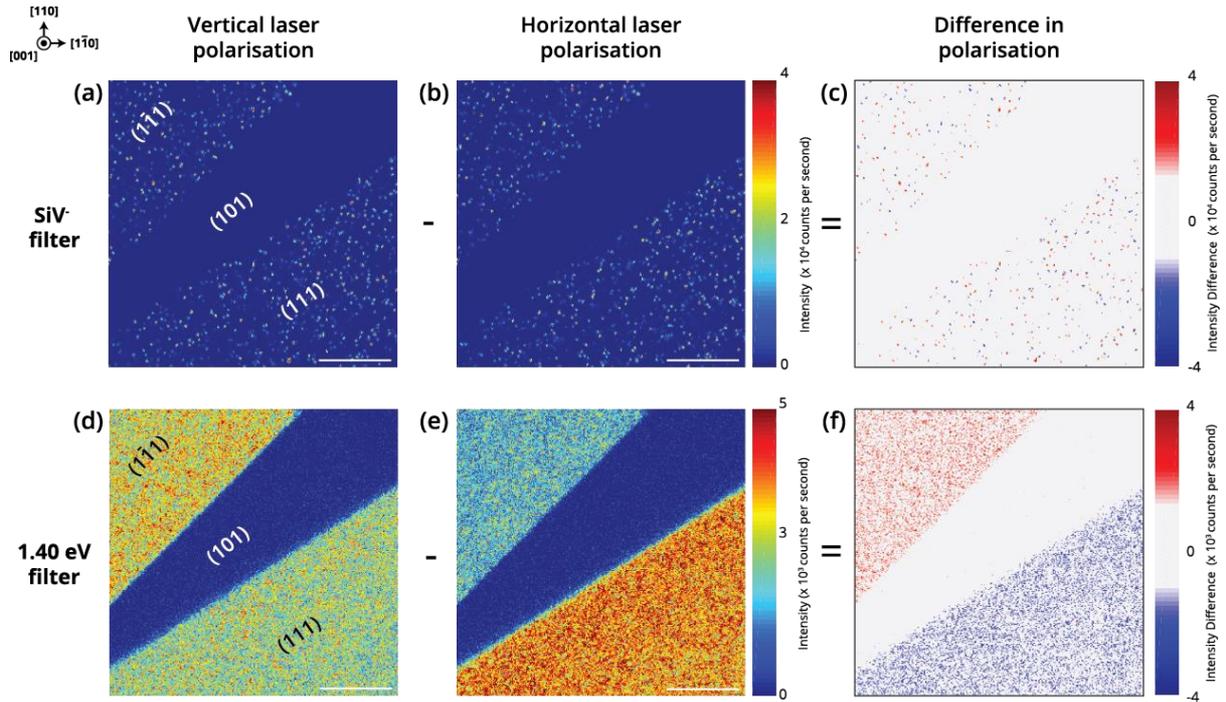

*Figure 7: The effect of laser polarisation (vertical or horizontal) on both SiV⁻, (a) and (b) respectively, and the 1.40 eV defect, (d) and (e) respectively. By taking the difference in intensity due to laser polarisation it is clear to see in (c) that SiV⁻ does not exhibit any preferential orientation, whereas there is a clear correlation in (f) for the 1.40 eV defect to indicate this defect could have grown in preferentially. (Scale bar; 10 μm).*

## 4 Discussion

### 4.1 Growth sectors and the nitrogen and boron incorporation

HPHT diamond crystals grown from solutions consist of (growth) sectors grown in discrete directions defined by the normal of the crystal faces involved. The final crystal exhibits only those faces which have low growth velocities since those with higher growth rates grow out and the external morphology is dominated by the slower growing sectors. It is well known that in HPHT diamond the incorporation probability for nitrogen and boron varies between growth sectors and that the actual uptake of impurities is influenced by the details of the growth conditions at each growth interface of the growing crystal.[7] Growth sectors are separated by growth-sector boundaries which are internal surfaces over which the edges between neighbouring sectors have passed during growth.[49] The local direction of a growth-sector boundary depends on the relative growth velocity of the neighbouring faces. If the relative velocity is constant the boundary is planar, but if not the boundary can be an irregular internal surface. In HPHT diamond the fast growing {110} sectors often have irregular surfaces with neighbouring slower growing {100} and {111} sectors.[3] The growth-sector boundaries and the surrounding regions may be indistinguishable from the bulk sectors in terms of defect and impurity incorporation, however there may be increased local impurity and defect incorporation when growth layers on neighbouring faces meet.

It is shown that the incorporation of boron into the {001} sector of the HPHT diamond studied is less than 1 ppb (Table 1). The large variation of boron concentration between growth sectors is consistent with published data[7,50,51] where it was shown that the boron is incorporated in the greatest concentrations in the {111} sectors, then in the {110} sectors, and in lower concentrations in {113}, {115} and {001} sectors. Our results and those of Klepikov et al[50] suggest that the boron concentration is lowest in the {001} sector, and very much less ($> \times 100$) than in the {111} sector.

We note that the $N_s^0$ concentration is well below the detection limit of Secondary Ion Mass Spectroscopy. Since the concentration of $N_s^0$ determined by EPR is an average over the whole sample it is difficult to know for certain the $N_s^0$ concentration in an individual growth sector. However, from the boron concentrations present in this sample (Table 1) we make the assumption that the nitrogen is mostly compensated (e.g. predominately in the form of $N_s^+$ which is not detected by EPR) in the {111} growth sectors, of which contributes 69% of the sample, but then only a fraction is compensated in the remaining sectors. By approximating the volume occupied by each growth sector, and allowing the relative growth sector dependence of the $N_s^0$ concentration to follow that determined by Burns *et al*[7], the measured EPR signal for $N_s^0$ is achieved if the uncompensated concentration in the {001} growth sector is ~50 ppb.

Further speculation is unwarranted; however, it is clear that at least in the {001} growth sector the concentration of boron impurities in HPHT diamond can be reduced to less than 1 ppb. Even if the $N_s^0$ concentration were as high as ~100 ppb we would expect the spin decoherence of any defects introduced to be limited by interaction with the natural abundance $^{13}$C. Given the low background boron contamination in the {001} growth sector we are optimistic that NV$^-$ defects introduced into this material by ion implantation or femto-second laser writing[52] would have very attractive properties that would not be compromised by grown-in strain given the very low concentration of dislocations. If the $N_s^0$ contamination of {001} growth sectors could be reduced to less than 1 ppb then this HPHT material may even outperform "quantum grade" CVD diamond in certain applications.

*4.2 Significance of NV decoration at the {111} to {113} growth sector boundary.*

The orientation of a growth sector boundary is dependent on the relative growth velocity of the two neighbouring crystal faces involved. The boundary is straight when the relative growth rate is constant, and curved when it changes due to fluctuations in the growth conditions. The relative growth rate can be determined if the growth-sector boundary is visible and the growth direction of the adjacent crystal faces can be identified.[49] The region shown in Figure 4a is particularly complicated as there is growth in several different directions. Taking a linear growth rate in {111} and {113} directions for HPHT synthesis at 1500 °C of ~50 µm/hour[53] would suggest that since the decoration of the growth sector boundary appears straight in Figure 4, the different sector growth rates were stable over tens of minutes.

The cross product of the vectors normal to the intersecting planes gives a vector in the plane of the growth sector boundary, and the direction of travel of the boundary is determined from the growth velocities of the two planes. From these two vectors (direction of travel of the boundary and cross product of two interesting planes) the plane of the growth sector boundary can be determined. In Figure 3, location A, where the observation direction is along ~ [001], the boundary between (113) - (111) and ($1\bar{1}3$) - ($1\bar{1}1$) growth sectors will be independent of the relative growth velocities and orientated along $[1\bar{1}0]$ (horizontal) and $[110]$ (vertical) directions, respectively.

The decorated boundary between ($1\bar{1}3$) - (111) indicates $v_{1\bar{1}3}/v_{111} \approx 1.13$, where $v_{lmn}$ is the linear growth velocity. In Figure 4b, the angle between the decorated boundary and the vertical [001] suggests that $v_{1\bar{1}3}/v_{1\bar{1}1} \approx 0.87$ and $v_{1\bar{1}3}/v_{111} \approx 1.14$ with the latter value in good agreement with the result for the [001] direction of observation. From Figure 4c, even though the boundaries are not decorated, the sector dependent defect incorporation allows us to estimate that $v_{101}/v_{1\bar{1}1} \approx 1.2$ and $v_{101}/v_{111} \approx 0.94$, but in the latter case the uncertainty is very large. Since there is no knowledge of where the plate was cut from the grown crystal, or its final morphology, it would be unwise to further analyse growth rate data noting this study represents only a small time window during growth.

It is interesting to consider the significance of the fact that only the {113} - {111} growth sector interface is decorated with NV⁻ defects. This transition zone between the {113} and {111} planes is seen in the visible CL image (Figure 2c) as a band of higher intensity emission, where the confocal maps correlate the width of these growth sector interfaces to the width of NV$^-$ decoration. The minimum angle change that can occur between two neighbouring growth sectors is outlined in Table 2. It is possible that is not energetically favourable for the crystal to grow with a direct transition of {113} to {111} (or {111} to {110}) and must do so via small steps with higher order crystallographic planes with subsequent increased incorporation of nitrogen and NV$^-$ defects at the {113} to {111} boundary. However, kinetic and impurity absorption effects may well influence the nature of the transition and it will require further work to determine why only one growth sector boundary is decorated with NV$^-$ defects.

*Table 2: The minimum angle, in degrees, that can occur between two neighbouring growth sectors.*

|  | Growth Sector Boundary | | | |
| --- | --- | --- | --- | --- |
|  | {111} - {113} | {111} - {011} | {113} - {115} | {115} - {001} |
| **Minimum angle** | 29.5° | 35.3° | 9.4° | 15.8° |

### *4.3 The lack of NV preferential orientation*

Preferential orientation of grown-in defects has been observed in both CVD[34,54]- and HPHT[48,55]-grown diamond, with some theoretical work carried out to investigate the surface kinetics and chemistry required to produce defect preferential orientation.[56] In the specific case of the NV$^{-/0}$ defect, there are numerous reports of preferential alignment during CVD growth on ⟨110⟩, ⟨113⟩ and ⟨111⟩ oriented substrates.[22,23,34] Destruction of preferential orientation has also been demonstrated by annealing samples at temperatures high enough to allow the defect in question to re-orientate.[57]

There are two plausible explanations for the lack of preferential orientation of the NV centres:
- the NV defect did not grow in as a unit but was formed by vacancies migrating to substitutional nitrogen defects to produce defects in all possible orientations.
- or, preferential orientation did occur during growth, but the growth temperature was high enough to enable reorientation of NV centres during the growth run without all the defects annealing out.

The sample studied has spent time at elevated temperatures in the growth capsule. Typical catalyst-solvent HPHT diamond growth temperatures are 1350 - 1600 °C.[51] The degree to which a defect "X" may undergo re-orientation can be modelled by first-order chemical kinetics. Using an activation energy $E_A \approx 4.0$ eV and attempt frequency $\nu \approx 10^{11}$ s$^{-1}$ for NV,[58] we find that no appreciable preferential orientation remains after only a few minutes (seconds) at 1350 °C (1500 °C). It is therefore reasonable to expect that any initial preferential orientation of NV would have annealed back to equilibrium populations during the time the sample spent at elevated temperatures even if the material studied was grown towards the end of the growth run. Using the values published by Pinto *et al*[59] for the activation energy of NV migration and attempt frequency, any NV preferential orientation would be lost in under 1 minute at 1500 °C. Thus the lack of observed preferential orientation of NV centres is consistent with our understanding of the thermal re-orientation of this defect. The one report of preferentially aligned NV centres produced in {111} growth sectors during HPHT growth[44] (orientation parallel to growth direction is missing) is at odds with our understanding.

The NV defect reorients locally before it dissociates/migrates and interacts with other defects and impurities. It is well known that as NV anneals out the di-nitrogen vacancy defect $N_2V$ is produced presumably by the reaction $NV+N_s \rightarrow N_2V$.[58,60,61] $N_2V^0$ gives rise to the H3 absorption/emission (ZPL 503 nm) and $N_2V^-$ to the H2 absorption/emission (ZPL 982 nm) and a characteristic EPR spectrum.[62] No H3 emission was detected using 488 nm excitation in any region of the sample using confocal PL. This suggests an upper limit for the growth temperature of 1500 °C since at this temperature, H3 could be produced.[58]

### 4.4 Preferential orientation of SiV$^-$ and the 1.40 eV defect

It is not surprising that preferential orientation of the 1.40 eV defect is seen in the {111} growth sectors since this has been previously reported.[57,63] The sector-dependent contrast follows an approximate 3:1 intensity contrast ratio for the 1.4 eV emission which would be expected for 100% preferential alignment of the symmetry axis of a trigonal defect (A ↔ E optical transition) along the growth direction of each of the respective {111} sectors and hence is clear evidence that this defect is indeed preferentially orientated. This HPHT diamond was synthesised in a Co-Fe-C system. Previous reports using the Co-Fe-C system with additives of titanium[64] or zirconium[65] to reduce nitrogen incorporation in the grown diamond have shown that high-quality diamond crystals can be produced. Thus the observation of the 1.40 eV defect in the {111} growth sectors of the diamond leads us to suppose that the growth system was contaminated with nickel. We note that if the concentration of the 1.40 eV defect were as low as ~ 0.1 ppb there would still be ~20 defects in the confocal PL volume. Thus if the nickel is readily incorporated the contamination of the growth system could be at trace amounts.

The absence of any preferential orientation of the SiV⁻ centres is unexpected. Preferential alignment has been previously predicted by first-principle calculations[66] and observed experimentally in CVD diamond doped with silicon.[44,54] In CVD grown diamond it has been suggested that like NV⁻, the SiV⁻ defect is grown in as a unit.[54] With regards to the reorientation of the SiV⁻ centres during growth, little is known about the migration energies involved, but even at temperatures as high as 1900 °C the SiV⁻ fluorescence intensity does not reduce significantly,[67] with preferential orientation of SiV⁰ persisting in CVD diamond annealed up to 2000 °C,[54] indicating that the defect is stable well above the anticipated HPHT growth temperature. The SiV⁻ defects observed in the {111} growth sectors of the HPHT diamond do not appear to be preferentially orientated, although we note a previous report suggesting that the orientation parallel to the [111] growth direction was missing in the HPHT diamond studied.[44] Further studies on the incorporation mechanism of SiV⁻ during HPHT synthesis is required with regards to the lack of preferential orientation in {111} sectors.

## 5   Conclusions

HPHT-synthesised diamond from NDT has been studied with confocal PL microscopy. In this material we find three species of point defects; NV⁻, SiV⁻, and nickel-related emission centred on 884 nm which we refer to in this paper as the 1.40 eV defect. Confocal imaging reveals sector interfaces between {113} and {111} are decorated with NV⁻ centres. We note that this decoration has now also been seen in type II HPHT diamonds produced by other laboratories; in all the high purity HPHT material studied to date decoration of {113} and {111} boundaries with NV⁻ centres is always observed. This decoration allows for determination of the relative growth rates of the neighbouring {113} and {111} growth sectors. The bulk {111} sectors in the sample studied here exhibit emission from only SiV⁻ and 1.40 eV defects when excited with 532 nm light. There is no evidence to suggest preferential orientation for either the NV⁻ or SiV⁻ defects although it is evident that 1.40 eV defect is grown-in with the trigonal axis of the defect aligned with the [111] growth direction, in line with previously published results. At the HPHT growth temperature, the NV⁻ defect is expected to rapidly re-orientate so no preferential orientation is expected even if the defect grew in as a unit on for example a {111} growth surface but a further understand is required for the incorporation of SiV$^-$.

The high crystalline quality of this material coupled with its high-purity, resulting in a low point defect density, reflects the recent strides made in HPHT diamond synthesis methods. Future work will focus on investigating the native NV centres for their suitability for quantum applications and how this material compares to HPHT diamond grown in different processes.

**Acknowledgements**

This work was supported by the EPSRC Centre for Doctoral Training in Diamond Science and Technology (Grant No. EP/L015315/1) and funding from the Gemological Institute of America for a PhD studentship for PLD. BLG and MEN acknowledge funding from NICOP (Grant No. N62909-16-1-2111-P00002 - Towards a Picotesla DC Diamond Magnetometer) and NQIT (Grant No. EP/M013243/1 - UK Quantum Technology Hub: NQIT - Networked Quantum Information Technologies). The authors would like to thank A.V. Koliadin, N.T. Khikhinashvili, R.V. Isakov and Y.A. Loguinov at New Diamond Technology for the growth of this sample, and D. J. F. Evans at De Beers Technologies UK for providing assistance with cathodoluminescence imaging and spectroscopy.